\documentclass[preprint,showkeys,showpacs,prl,onecolumn,superbib]{revtex4}%
\usepackage{amsfonts}
\usepackage{amsmath}
\usepackage{amssymb}
\usepackage{graphicx}%
\newtheorem{theorem}{Theorem}

\newtheorem{corollary}[theorem]{Corollary}

\newtheorem{definition}{Definition}

\ifx\pdfoutput\relax\let\pdfoutput=\undefined\fi
\newcount\msipdfoutput
\ifx\pdfoutput\undefined\else
\ifcase\pdfoutput\else
\ifx\paperwidth\undefined\else
\ifdim\paperheight=0pt\relax\else\pdfpageheight\paperheight\fi
\ifdim\paperwidth=0pt\relax\else\pdfpagewidth\paperwidth\fi
\fi\fi\fi
\begin{document}
\title[New Bell inequalities]{ New Bell inequalities for the singlet state: Going beyond the Grothendieck bound.}
\author{Itamar Pitowsky}
\affiliation{The Hebrew University, Mount Scopus, Jerusalem 91905, Israel.}

\begin{abstract}
Contemporary versions of Bell's argument against local hidden variable (LHV)
theories are based on the Clauser Horne Shimony and Holt (CHSH) inequality,
and various attempts to generalize it. The amount of violation of these
inequalities cannot exceed the bound set by the Grothendieck constants.
However, if we go back to the original derivation by Bell, and use the perfect
anticorrelation embodied in the singlet spin state, we can go beyond these
bounds. In this paper we derive two-particle Bell inequalities for traceless
two-outcome observables, whose violation in the singlet spin state go beyond
the Grothendieck constants both for the two and three dimensional cases.
Moreover, creating a higher dimensional analog of perfect correlations, and
applying a recent result of Alon and his associates (\emph{Invent. Math. 163}
499 (2006)) we prove that there are two-particle Bell inequalities for
traceless two-outcome observables whose violation increases to infinity as the
dimension and number of measurements grow. Technically these result are
possible because perfect correlations (or anticorrelations) allow us to
transport the indices of the inequality from the edges of a bipartite graph to
those of the complete graph. Finally, it is shown how to apply these results
to mixed Werner states, provided that the noise does not exceed 20\%.

\ \ \ \ \ \ \ \ \ \ \ \ \ \ \ \ \ \ \ \ \ \ \ \ \ \ \ \ \ \ \ \ \ \ \ \ \ \ \ \ \ \ \ \ \ \ \ \ \ \ \ \ \ \ \ 

\end{abstract}
\startpage{1}
\maketitle
\published{\emph{Journal of Mathematical Physics 49}, 012101 (2008).}

\section{Introduction}

The impossibility of reproducing all correlations observed in composite
quantum systems using local hidden variables (LHV) was proven in 1964 by Bell.
In his work~\cite{1}, Bell showed that local models satisfy the Bell
inequality, but there are measurements on the singlet quantum state that
violate it. Contemporary versions of the argument are based on the
Clauser-Horne-Shimony-Holt (CHSH) inequality \cite{2}, and not the original
inequality used by Bell. There is a very good reason for that. While Bell's
argument applied only to the singlet state, the CHSH inequality is violated by
all pure entangled states \cite{3}, and also by certain mixtures, provided
they are not too noisy. From an experimental point of view this is a
substantial difference. In recent years many generalizations of CHSH\ have
been introduced (see Refs \cite{4}, \cite{5} for details).

However, if we are interested in the question what is the maximal possible
violation of any Bell inequality? It is profitable to go back to the original
version of Bell (or its simplification by Wigner \cite{6}). Bell is using the
perfect anticorrelation embodied in the singlet spin state. This means that
when the spin measurements performed on both sides are along the same
direction the outcomes are always opposite. A LHV theory that attempts to
reproduce an experiment on the singlet state is thus further constrained.

In this paper, I shall show how this allows to increase the violation in the
singlet spin state beyond the CHSH inequality and its generalizations. The
possible violations of the latter are bounded by the Grothendieck constants
\cite{7}. We shall see in the third section that in both the two and three
dimensional cases, we can get a violations beyond the corresponding
Grothendieck constants. Moreover, using the theorem of Tsirelson \cite{8}, we
can create an analog of perfectly correlated states in any dimension $n$
(section 4). Applying a recent result of Alon \emph{et.al.} \cite{9}, I prove
that there are Bell inequalities whose violation \emph{increases to infinity}
with $n$. This should be compared with the traditional approach, based on
extensions of the CHSH, where the violation in any dimension cannot increase
beyond a finite bound, the Grothendieck constant $K_{G}$. Technically, this
outcome is possible because perfect correlations (or anticorrelations) allow
us to transport the indices of the inequality from the edges of a bipartite
graph to those of the complete graph.

The present result adds to a recent proof that the violation of three party
Bell inequalities can grow unbounded \cite{10}. Both these results strengthen
the insight of Mermin \cite{11} that macroscopic objects do not necessarily
behave in an approximately classical way. Mermin considered $k$ spin
${\frac12}$
particles in the symmetric GHZ state, derived a Bell inequality involving two
measurements per particle each with two possible outcomes, and showed that the
violation of the inequality grows exponentially with $k$. A natural question
is whether we can get unbounded violations with a fixed number of high spin
particles. In \cite{10} the three particles case is settled in the
affirmative. In the two particles case the Grothendieck bound seems to imply
that at least for the two outcomes zero trace observables the answer is
negative. In this paper it is shown that this problem can be bypassed if the
state is symmetric.

Furthermore, in section 5 these results are extended to mixed Werner states
\cite{12}, provided the correlations are sufficiently close to $-1$ (or to
$+1$). This shows that Bell's original approach can be applied to noisy cases,
not just the pure singlet. Here, we do not get as good results as those
obtained by CHSH. However, it turns out that the new inequalities are robust
against less than $20\%$ noise. Moreover, given a fixed amount of noise in
this range, the violation still goes to infinity with $n$.

\section{Local hidden variables, Bell inequalities and symmetries}

Let Alice and Bob share a singlet state%
\begin{equation}
\,|\psi\rangle=\frac{1}{\sqrt{2}}\,(\,|+\rangle\,|-\rangle\,-|-\rangle
\,|+\rangle)\, \label{1}%
\end{equation}
Consider a spin measurement along the $\mathbf{x}$ direction performed by
Alice, and along the $\mathbf{y}$ direction by Bob (all Bob's parameters will
be denoted with $y$'s Alice's with $x$'s). The operator corresponding to both
measurement is $S_{\mathbf{x}}\otimes S_{\mathbf{y}}$. The expectation of this
operator in the singlet state is $\langle\psi|S_{\mathbf{x}}\otimes
S_{\mathbf{y}}|\psi\rangle=-\mathbf{x\cdot y}$, and in case $\mathbf{x=y}$ the
expectation is $-1$. If Alice is choosing her measurements to be in either the
$\mathbf{x}_{1}$ or $\mathbf{x}_{2}$ directions, and Bob is choosing between
$\mathbf{y}_{1}$ and $\mathbf{y}_{2}$, there are four possible arrangements
with the expectations $\langle\psi|S_{\mathbf{x}_{i}}\otimes S_{\mathbf{y}%
_{j}}|\psi\rangle=-\mathbf{x}_{i}\mathbf{\cdot y}_{j}$ , $i=1,2$ and $j=1,2$.

A (deterministic) hidden variable theory associates with the physical system a
parameter $\lambda$ whose value determines the outcome of every such
experiment. The hidden variable theory is \emph{local} if the value that Alice
is measuring does not depend on the direction chosen on Bob's side, and vice
versa. Let $X_{i}(\lambda)$ be the value that Alice is obtaining when the
hidden variable has the value $\lambda$ and the direction on her side is
$\mathbf{x}_{i}$. By definition $X_{i}(\lambda)=\pm1$. Similarly let
$Y_{j}(\lambda)$ be the value measured on Bob's side. Since our LHV is assumed
to recover the quantum correlations we expect that when we average the value
of $X_{i}(\lambda)Y_{j}(\lambda)$ over the space of hidden variables we shall
recover the quantum correlations, in other words%
\begin{equation}
E(X_{i}Y_{j})\equiv\int X_{i}(\lambda)Y_{j}(\lambda)d\mu(\lambda)=\langle
\psi|S_{\mathbf{x}_{i}}\otimes S_{\mathbf{y}_{j}}|\psi\rangle=-\mathbf{x}%
_{i}\mathbf{\cdot y}_{j}.\label{2}%
\end{equation}
Here, $\mu$ is the probability measure on the space of hidden variable.
However, LHV theories do not always exist. To see that consider the inequality%
\begin{equation}
\frac{1}{2}X_{1}Y_{1}+\frac{1}{2}X_{1}Y_{2}+\frac{1}{2}X_{2}Y_{1}-\frac{1}%
{2}X_{2}Y_{2}\leq1,\label{3}%
\end{equation}
which is satisfied by every choice of values $\pm1$ to the $X_{i}$'s and
$Y_{j}$'s. This means that the expectations also satisfy%
\begin{equation}
\frac{1}{2}E(X_{1}Y_{1})+\frac{1}{2}E(X_{1}Y_{2})+\frac{1}{2}(X_{2}%
Y_{1})-\frac{1}{2}E(X_{2}Y_{2})\leq1.\label{4}%
\end{equation}
Now, choose $\mathbf{x}_{1}=(-\frac{\sqrt{2}}{2},\frac{\sqrt{2}}{2},0)$,
$\mathbf{x}_{2}=(\frac{\sqrt{2}}{2},\frac{\sqrt{2}}{2},0)$, $\mathbf{y}%
_{1}=(0,-1,0)$, $\mathbf{y}_{2}=(1,0,0)$ and the quantum expectation values in
(\ref{2}) satisfies%

\begin{equation}
-\frac{1}{2}\mathbf{x}_{1}\mathbf{\cdot y}_{1}-\frac{1}{2}\mathbf{x}%
_{1}\mathbf{\cdot y}_{2}-\frac{1}{2}\mathbf{x}_{2}\mathbf{\cdot y}_{1}%
+\frac{1}{2}\mathbf{x}_{2}\mathbf{\cdot y}_{2}=\sqrt{2}. \label{5}%
\end{equation}

This result can be given a geometric interpretation. In the four dimensional
real space $\mathbb{R}^{4}$ consider the convex hull of the $16$ vectors of
the form $(X_{1}Y_{1},X_{1}Y_{2},X_{2}Y_{1},X_{2}Y_{2})$ where $X_{i}%
,Y_{j}=\pm1$. This is the Bell polytope $Bell(2,2)$, and the inequality
(\ref{3}) is one of its non trivial facets. We normalize this inequality so
the constant on the right hand side of is $1$. In this way the amount of
violation of inequality (\ref{3}) given in (\ref{5}) indicates how far the
vector $(-\mathbf{x}_{1}\mathbf{\cdot y}_{1},-\mathbf{x}_{1}\mathbf{\cdot
y}_{2},-\mathbf{x}_{2}\mathbf{\cdot y}_{1},-\mathbf{x}_{2}\mathbf{\cdot y}%
_{2})=(\frac{\sqrt{2}}{2},\frac{\sqrt{2}}{2},\frac{\sqrt{2}}{2},-\frac
{\sqrt{2}}{2})$ is from the boundary of $Bell(2,2)$.

It is interesting that we can increase this violation in the case of the
singlet. To see that we just repeat, more or less, Bell's original argument
\cite{1}. Consider the following inequality which refers only to Alice's
hidden variables%
\begin{equation}
-X_{1}X_{2}-X_{1}X_{3}-X_{2}X_{3}\leq1.\label{6}%
\end{equation}
Assume that the $X_{i}(\lambda)$, $i=1,2,3$ are the LHV values of the results
of Alice's measurements in the direction $\mathbf{x}_{i}$. Suppose that Bob
chooses his measurements to be in exactly the same directions. If $Y_{1}%
,Y_{2},Y_{3}$ are the hidden variables corresponding to his results then
quantum mechanics predicts that $E(X_{i}Y_{i})=-\mathbf{x}_{i}\cdot
\mathbf{x}_{i}=-1$, $i=1,2,3$. Since the $X_{i}$, $Y_{j}$ can have only the
values $\pm1$, we must conclude that $X_{i}(\lambda)=-Y_{i}(\lambda)$ for
almost all $\lambda$ (with respect to the probability measure $\mu$ on the
space of hidden variables), and therefore,
\begin{equation}
E(X_{i}X_{j})=-E(X_{i}Y_{j})=\mathbf{x}_{i}\cdot\mathbf{x}_{j}.\label{7}%
\end{equation}
From the trivial inequality (\ref{6}), we have on the other hand:
$-E(X_{1}X_{2})-E(X_{1}X_{3})-E(X_{2}X_{3})\leq1$. However, if we take
$\mathbf{x}_{1},\mathbf{x}_{2},\mathbf{x}_{3}$ to be in the same plane and
$120^{\circ}$ apart we get from (\ref{7})
\begin{equation}
-\mathbf{x}_{1}\mathbf{\cdot x}_{2}-\mathbf{x}_{1}\mathbf{\cdot x}%
_{3}-\mathbf{x}_{2}\mathbf{\cdot x}_{3}=\frac{3}{2}>\sqrt{2}.\label{8}%
\end{equation}
Here too, we can associate a geometric picture with inequality (\ref{6}). In
$\mathbb{R}^{3}$ consider the set of $8$ vectors of the form $(X_{1}%
X_{2},X_{1}X_{3},X_{2}X_{3})$ with $X_{i}=\pm1$, their convex hull is a
polytope $Bell(3)$ and inequality (\ref{6}) represents a facet. Again, we
choose the constant on the right hand side of (\ref{6}) to be $1$, and
(\ref{8}) indicates how far is the vector $(\mathbf{x}_{1}\mathbf{\cdot x}%
_{2},\mathbf{x}_{1}\mathbf{\cdot x}_{3},\mathbf{x}_{2}\mathbf{\cdot x}%
_{3})=(-\frac{1}{2},-\frac{1}{2},-\frac{1}{2})$ from the boundary of
$Bell(3)$. \ More on these polytopes and their generalizations in the appendix A.

To test this violation experimentally we take pairs from a source in the state
(\ref{1}). In each run we measure the spin of the left particle along one of
the directions $\mathbf{x}_{1}\mathbf{,x}_{2}$, or $\mathbf{x}_{3},$ and the
spin of the right particle along one of these same directions. From the
subensemble in which the same direction is chosen on both sides we can infer
the symmetry, and from the other subensembles we get the correlations.
Finally, we substitute these values into (\ref{8}) to see if a violation
occurs. Hence, each component of (\ref{8}) can be tested, and the amount of
violation beyond the classical case is a physical parameter. The same applies
to the other inequalities below.

Note that so far we dealt only with directions $\mathbf{x}_{i}$ which lie in a
two dimensional plane. In the next section, we shall see how to use the same
method to get a violation in three dimensions above the known upper bounds.
This is achieved by the application of a new Bell inequality. More generally,
consider the possible extensions of the CHSH method. Define the number
$K_{G}(n)$- called the Grothendieck constant of order $n$- to be the least
positive real number such that the inequality%
\begin{equation}
\left\vert \sum_{i=1}^{m}\sum_{j=1}^{m}a_{ij}\,\mathbf{x}_{i}\cdot
\mathbf{y}_{j}\right\vert \leq K_{G}(n)\underset{X_{i},Y_{j}=\pm1}{\sup
}\left\vert \sum_{i=1}^{m}\sum_{j=1}^{m}a_{ij}\,X_{i}Y_{j}\right\vert
,\label{9}%
\end{equation}
is satisfied for every natural number $m$, every choice of real numbers
$a_{ij}$, $1\leq i\leq m$, $1\leq j\leq m$, and every choice of unit vectors
$\mathbf{x}_{1},...,\mathbf{x}_{m},\mathbf{y}_{1},...,\mathbf{y}_{m}%
\in\mathbb{R}^{n}$. Grothendieck \cite{13} introduced these constants and
proved that $K_{G}=\underset{n\rightarrow\infty}{\lim}K_{G}(n)$ is finite.
Tsirelson \cite{8} made the connection with Bell inequalities. It is known
that $K_{G}(2)=\sqrt{2}$, so (\ref{5}) is the best possible result when all
the directions are in the same plane. It is not known whether this is also the
best result when we let the directions $\mathbf{x}_{i}$, $\mathbf{y}_{j}$ vary
in the three dimensional space. In this case the known bounds are $\sqrt
{2}\leq K_{G}(3)\leq1.5163$. It is also known that the limit $K_{G}$ satisfies
$1.6770\leq K_{G}\leq\pi/(2\log(1+\sqrt{2}))=1.7822$ (see \cite{7} for
details). Although the last lower bound is better than $\sqrt{2}$, it is hard
to come up with a concrete inequality which will do better than CHSH, but some
examples are known \cite{14}. In general, finding new Bell inequalities, let
alone all of them, is a computationally hard problem \cite{15} (more on this
in the Appendix A). However, this does not mean that one cannot derive all the
inequalities when the number of measurements per site is small, or more
generally, in special infinite cases. An important example of the latter is
the set of all inequalities for $k$ parties with two dichotomic observable per
site \cite{16}.

However, we have already seen how we can get above $\sqrt{2}$ when we are
using perfect correlations. In the next section we shall see that our method
leads to a violation above the upper bound for $K_{G}(3)$ in the three
dimensional case. The reason why all this is possible is the following: In the
CHSH case (\ref{2}) the sum is taken over the edges of the bipartite graph
$K_{2,2}$, and in the higher dimensional case in equation (\ref{9}) over
$K_{m,m}$. However, in inequality (\ref{6} ) the sum is taken over the edges
of the complete graph on $3$ vertices $K_{3}$, and this makes the difference.
Indeed, let $A(n)$ be the least real number such that%
\begin{equation}
\sum_{1\leq i<j\leq n}a_{ij}\,\mathbf{x}_{i}\cdot\mathbf{x}_{j}\leq
A(n)\underset{X_{i}\ =\pm1}{\sup}\left(  \sum_{1\leq i<j\leq n}a_{ij}%
\,X_{i}X_{j}\right)  \label{10}%
\end{equation}
for every choice of reals $a_{ij}$, $1\leq i<j\leq n$, and every
$\mathbf{x}_{1},...,\mathbf{x}_{n}\in\mathbb{R}^{n}$. As $n$ grows $A(n)$
becomes unbounded; in a recent paper Alon \emph{et.al.} \cite{9} extended
Grothendieck's work to the complete graph and proved that $A(n)=\Omega(\log
n).$ In section 3, I will note how this theorem is relevant to quantum mechanics.

Finally in section 4 I show how to extend the method to cases where the
correlations are not perfect, for example in some mixed Werner states.
Although the inequalities are provably robust against quite a bit of noise,
the method so far did not lead to results as good as CHSH. Some open questions
regarding this point are indicated.

\section{The Clique-Web inequality}

\begin{definition}
Let $p,q$, and, $r$ be three integers such that $q\geq2$ and $p-q=2r+1$. The
\emph{web }$W_{p}^{r}$ is the graph whose set of vertices is $V_{p}%
=\{1,2,...,p\}$ and set of edges is $\{i,i\nolinebreak+\nolinebreak
r\nolinebreak+\nolinebreak1\},...,\{i,i+r+q\}\ ;\ i=1,2,...,p,\ $%
addition\ is$\ \operatorname{mod}p.$
\end{definition}

Figure 1 shows the web $W_{8}^{2}$; the similarity with the spider web is the
source of the name.

\begin{center}
Figure 1: the web $W_{8}^{2}$
\end{center}

Alon \cite{17} identified the maximal cuts in $W_{p}^{r}.$ Following his
result Deza and Laurent were able to find new facets of the cut polytope
\cite{18}. The facet inequalities of the cut polytope are very closely related
to Bell inequalities (see \cite{18}, more on that in the Appendix B). From
their result it is easy to derive the following:

\begin{theorem}
Let $p,q$, and, $r$ be three integers such that $q\geq2$ and $p-q=2r+1$. Then
the following inequality is satisfied for all $X_{1},...,X_{p},Z_{1}%
,...,Z_{q}\in\{-1,1\}$
\begin{equation}
\sum_{i=1}^{p}\sum_{j=1}^{q}X_{i}Z_{j}-\sum_{\{i,j\}\in W_{p}^{r}}X_{i}%
X_{j}-\sum_{1\leq i<j\leq q}Z_{i}Z_{j}\leq q(r+1). \label{11}%
\end{equation}

\end{theorem}

The inequality involves $p+q$ variables $X_{i}$ and $Z_{j}$. They can be taken
as the random variables in a LHV theory, representing the outcomes of Alice's
spin measurements along the $p+q$ directions $\mathbf{x}_{i}$ and
$\mathbf{z}_{j}$ in $\mathbb{R}^{3}$. Assuming, as before, that Bob is
measuring the spin along the same directions as Alice, and taking the perfect
anti-correlation in the singlet state into account, we obtain a contradiction
with the LHV model when
\begin{equation}
V_{p}^{r}=\frac{1}{q(r+1)}\left[  \sum_{i=1}^{p}\sum_{j=1}^{q}\mathbf{x}%
_{i}\cdot\mathbf{z}_{j}-\sum_{\{i,j\}\in W_{p}^{r}}\mathbf{x}_{i}%
\cdot\mathbf{x}_{j}-\sum_{1\leq i<j\leq q}\mathbf{z}_{i}\cdot\mathbf{z}%
_{j}\right]  >1. \label{12}%
\end{equation}

Let $p=12$ and $q=3$, $r=4$, and let $\mathbf{z}_{1}=\mathbf{z}_{2}%
=\mathbf{z}_{3}=(0,0,1)$. For $0<\theta<\frac{\pi}{2}$ let $\mathbf{x}%
_{1},\mathbf{x}_{2},...,\mathbf{x}_{12}$ all satisfy $\mathbf{x}_{i}%
\cdot\mathbf{z}_{j}=\cos\theta$, and be evenly spaced on the circle
$\{\mathbf{x\ };\mathbf{x\cdot z}_{j}=\nolinebreak\cos\theta\}$. The vectors
$\mathbf{x}_{i}$, $\mathbf{z}_{j}$ look like a bouquet of flowers (Figure 2).

\begin{center}
Figure 2: The 12-bouquet

\end{center}

We have $\mathbf{x}_{i}\cdot\mathbf{z}_{j}=\cos\theta$, $\mathbf{z}_{i}%
\cdot\mathbf{z}_{j}=1$, $\mathbf{x}_{i}\cdot\mathbf{x}_{i+6}=\cos2\theta$,
$\mathbf{x}_{i}\cdot\mathbf{x}_{i+5}=\mathbf{x}_{i}\cdot\mathbf{x}%
_{i+7}=(1-\,2\cos^{2}\frac{\pi}{12}\sin^{2}\theta)$, where the sum in the
indices is taken $\operatorname{mod}12$. Substituting these values into
(\ref{12}) we get%
\begin{equation}
V_{12}^{4}(\theta)=\frac{36\cos\theta-6\cos2\theta-12(1-\,2\cos^{2}\frac{\pi
}{12}\sin^{2}\theta)-3}{15},\label{13}%
\end{equation}
and $V_{12}^{4}(\theta)>1$ for a large range of $\theta$. In particular
$f(0.32477\pi)=\allowbreak1.\,\allowbreak520\,9$ is larger than the upper
bound for $K_{G}(3)$. In this way we can get sequences of independent
inequalities, with unbounded number of measurements, all violated by the
singlet state. One way to see that is to generalize the above example: we take
$p=2k+1$, $q=2$, and $r=k-1$, and define the $2k+1$-bouquet by taking
$\mathbf{z}_{1}=\mathbf{z}_{2}=(0,0,1)$, and $\mathbf{x}_{1},\mathbf{x}%
_{2},...,\mathbf{x}_{2k+1}$ be equally spaced and satisfy $\mathbf{x}_{i}%
\cdot\mathbf{z}_{j}=\cos\theta$, \ then%
\begin{equation}
V_{2k+1}^{k-1}(\theta)=\frac{(2k+1)[2\cos\theta-(1-\,2\cos^{2}\frac{\pi}%
{4k+2}\sin^{2}\theta)]-1}{2k},\label{14}%
\end{equation}
In this case, $V_{11}^{4}(\allowbreak0.330\,3\pi)=\allowbreak1.\,\allowbreak
516\,8$ is again larger than the upper bound for $K_{G}(3)$. Also,
$\underset{k\rightarrow\infty}{\lim}V_{2k+1}^{k-1}(\theta)=2\cos\theta
-\cos2\theta$ with the maximum $1.5$ obtained at $\theta=\frac{\pi}{3}$.

\section{An application of Tsirelson's theorem}

We can generalize the above argument using the following result \cite{8},

\begin{theorem}
\textbf{\ }The following conditions on an $n\times n$ matrix $(r_{ij})$ are equivalent:

\textbf{a}. There exists a finite dimensional Hilbert space $\mathcal{H}$,
Hermitian operators $A_{1,}A_{2},...A_{n}$, $B_{1,}B_{2},...B_{n}$, and a
state $W$ on $\mathcal{H}\otimes\mathcal{H}$ such that $spectrum[A_{i}%
]\subset\lbrack-1,1]$, $spectrum[B_{j}]\subset\lbrack-1,1]$ and $r_{ij}%
=tr[W(A_{i}\otimes B_{j})]$ for $i,j=1,2,...,n$.

\textbf{b}. The same as in 1, but with the additional conditions: $A_{i}%
^{2}=I$, $B_{j}^{2}=I$, $\ tr[W(A_{i}\otimes I)]=0$, $tr[W(I\otimes B_{j}%
)]=0$, $\ A_{i_{1}}A_{i_{2}}+A_{i_{2}}A_{i_{1}}$ is proportional to $I$ \ for
all $i_{1},i_{2}=1,2,...,n$, $\ B_{j_{1}}B_{j_{2}}+B_{j_{2}}B_{j_{1}}$ is
proportional to $I$ \ for all $j_{1},j_{2}=1,2,...,n$, and $\ \dim
\mathcal{H}\leq2^{[\frac{n+1}{2}]}$.

\textbf{c}. There exist unit vectors $\mathbf{x}_{1,}\mathbf{x}_{2}%
,...,\mathbf{x}_{n}$ and $\mathbf{y}_{1},\mathbf{y}_{2},...\mathbf{y}_{n}$ in
the $2n$-dimensional real space $\mathbb{R}^{2n}$ such that $r_{ij}%
=\mathbf{x}_{i}\cdot\mathbf{y}_{j}$.
\end{theorem}

From which we obtain the following.

\begin{corollary}
Given unit vectors $\mathbf{x}_{1,}\mathbf{x}_{2},...,\mathbf{x}_{n}%
\in\mathbb{R}^{n}$ there is a finite dimensional Hilbert space $\mathcal{H}$,
traceless Hermitian operators $A_{1,}A_{2},...A_{n}$, $B_{1,}B_{2},...B_{n}$
on $\mathcal{H}$ with spectrum $\pm1$, a state $W$ on $\mathcal{H}%
\otimes\mathcal{H}$ such that $tr[W(A_{i}\otimes I)]=0$, $tr[W(I\otimes
B_{j})]=0$, $tr[W(A_{i}\otimes B_{j})]=\mathbf{x}_{i}\cdot\mathbf{x}_{j}$. In
particular $tr[W(A_{i}\otimes B_{i})]=1$.
\end{corollary}

To see that just extend the $\mathbf{x}_{i}$'s to be $2n$- dimensional unit
vectors by adding zero coordinates, then define $\mathbf{y}_{j}=\mathbf{x}%
_{j}$ and apply Tsirelson's theorem. To make sure that $A_{i}$ and $B_{j}$ are
traceless we can always extend the Hilbert space $\mathcal{H}$ by adding
finitely many extra dimensions (without changing the support of $W$), and
making sure that $tr[A_{i}]=tr[B_{j}]=0$ by adding $\pm1$ to their spectrum.

Now, consider a sequence of measurements where Alice and Bob are sharing many
copies of the state $W$, Alice is measuring each one of the operators
$A_{1,}A_{2},...A_{n}$ several times, and Bob the operators $B_{1,}%
B_{2},...B_{n}$. The possible outcomes on each side are $\pm1$. A LHV model
for this experiment consists of an association of a random variable
$X_{i}(\lambda)$ with every operator $A_{i}$, and a random variable
$Y_{j}(\lambda)$ with $B_{j}$, such that $X_{i}(\lambda),Y_{j}(\lambda
)\in\{-1,1\}$ for every value of the hidden variable $\lambda$. To recover the
quantum correlation we require that $E(X_{i}Y_{j})=tr[W(A_{i}\otimes
B_{j})]=\mathbf{x}_{i}\cdot\mathbf{x}_{j}$. Since we have $E(X_{i}Y_{i})=1$ we
conclude that $X_{i}(\lambda)=Y_{i}(\lambda)$ for almost all $\lambda$ and all
$1\leq i\leq n$. This means that for the purpose of this LHV model $W$ behaves
very much like the singlet, only that now the $\mathbf{x}_{i}$'s are
directions in $\mathbb{R}^{n}$.

Charikar and Wirth \cite{19} proved that there is a universal constant $C>0$
such that%
\begin{equation}
\sum_{1\leq i<j\leq n}a_{ij}\,\mathbf{x}_{i}\cdot\mathbf{x}_{j}\leq C\log
n\ \underset{X_{i}=\pm1}{\sup}\left(  \sum_{1\leq i<j\leq n}a_{ij}\,X_{i}%
X_{j}\right)  \label{15}%
\end{equation}
for every real numbers $a_{ij}$, $1\leq i<j\leq n$ and all unit vectors
$\mathbf{x}_{1},\mathbf{x}_{2}...,\mathbf{x}_{n}\in\mathbb{R}^{n}$. More
important for our purpose, it was lately proved \cite{9} that \emph{this is
the best bound}. In other words \emph{there is a universal constant }%
$c>0$\emph{\ such that for each }$n$\emph{\ there are real numbers }$b_{ij}%
$\emph{, }$1\leq i<j\leq n$\emph{\ and unit vectors }$\mathbf{x}%
_{1,}\mathbf{x}_{2},...,\mathbf{x}_{n}\in\mathbb{R}^{n}$\emph{\ with}%
\begin{equation}
\sum_{1\leq i<j\leq n}b_{ij}\,\mathbf{x}_{i}\cdot\mathbf{x}_{j}>c\log
n\ \underset{X_{i}=\pm1}{\sup}\left(  \sum_{1\leq i<j\leq n}b_{ij}\,X_{i}%
X_{j}\right)  .\label{16}%
\end{equation}
Now choose the directions $\mathbf{x}_{i}$ that are guaranteed by this result
and the corresponding quantum system in Corrolary 3. Inequality (\ref{16})
then means that the results of Alice and Bob's measurements $tr[W(A_{i}\otimes
B_{j})]=\mathbf{x}_{i}\cdot\mathbf{x}_{j}$ violate local realism by an amount
which is unbounded, and increases to infinity with the dimension and with the
number of measurements.

As noted, this is possible because our LHV theory is constrained to satisfy
the condition of perfect correlations. Below we shall see how this condition
can be somewhat relaxed.

\section{The case of Werner states}

Let $|\psi\rangle$ be the singlet in (\ref{1}) and consider the mixed state
introduced by Werner \cite{12}%
\begin{equation}
\rho_{\eta}=\eta\,|\psi\rangle\!\langle\psi|+\frac{1-\eta\,}{4}I.\label{17}%
\end{equation}
here $0<\eta<1$ and $I$ the unit operator on the four dimensional complex
Hilbert space. If $\eta<\frac{1}{3}$ the state $\rho_{\eta}$ is separable and
if $\eta>\frac{1}{\sqrt{2}}$, it violates the CHSH inequality (\ref{2}). The
interesting feature of this state is that there are values of $\eta$ for which
$\rho_{\eta}$, although not separable, still admits LHV models of various
kinds (for details see \cite{4,5}). My purpose here is to show how Bell's
original argument can be extended to this case of imperfect correlations. This
does not improve the $\frac{1}{\sqrt{2}}$ bound given by CHSH, but better
inequalities may do just that.

We have $tr[\rho_{\eta}(S_{\mathbf{x}}\otimes S_{\mathbf{y}})]=-\eta
\mathbf{x\cdot y}$, and in particular $tr[\rho_{\eta}(S_{\mathbf{x}}\otimes
S_{\mathbf{x}})]=-\eta$. As before, let Alice measure the spin in three
directions $\mathbf{x}_{1,}\mathbf{x}_{2},\mathbf{x}_{3}$ and let Bob measure
in the spin in the same directions. Assume that $X_{i}(\lambda)=\pm1$ is the
value assigned to Alice's measurements in the $\mathbf{x}_{i}$ direction by a
LHV model, and likewise let $Y_{i}(\lambda)$ be Bob's value.%
\begin{equation}
E(X_{i}Y_{j})=-\eta\mathbf{x}_{i}\mathbf{\cdot x}_{j},\;\;E(X_{i}Y_{i}%
)=-\eta\label{18}%
\end{equation}
Again, let $\mu$ be the probability measure on the space of hidden variables.
Put $A_{+}^{j}=\{\lambda\ ;X_{j}(\lambda)=Y_{j}(\lambda)\}$ and $\ A_{-}%
^{j}=\{\lambda\ ;X_{j}(\lambda)=-Y_{j}(\lambda)\}$. Then $\mu(A_{+}^{j}%
)+\mu(A_{-}^{j})=1$, and by (\ref{18}): $\mu(A_{+}^{j})-\mu(A_{-}^{j}%
)=E(X_{j}Y_{j})=-\eta$, hence $\mu(A_{+}^{j})=\frac{1}{2}(1-\eta)$. Now, let
$i\neq j$ then%
\[
E(X_{i}X_{j})=\int_{A_{+}^{j}}X_{i}Y_{j}d\mu(\lambda)-\int_{A_{-}^{j}}%
X_{i}Y_{j}d\mu(\lambda)=2\int_{A_{+}^{j}}X_{i}Y_{j}d\mu(\lambda)-E(X_{i}%
Y_{j}),
\]
and therefore,%
\begin{equation}
-(1-\eta)-E(X_{i}Y_{j})\leq E(X_{i}X_{j})\leq(1-\eta)-E(X_{i}Y_{j}).\label{19}%
\end{equation}

Consider inequality (\ref{6}), only written in a slightly different way,%
\begin{equation}
-X_{1}X_{2}+X_{1}Y_{3}+X_{2}Y_{3}\leq1.\label{20}%
\end{equation}
Using (\ref{18}), and assuming the worst case in (\ref{19}) we conclude that
the following inequality must be satisfied by a LHV model%
\begin{equation}
E(X_{1}Y_{2})-(1-\eta)+E(X_{1}Y_{3})+E(X_{2}Y_{3})\leq1.\label{21}%
\end{equation}
Hence, a violation occurs when $-\eta\mathbf{x}_{1}\mathbf{\cdot x}%
_{2}-(1-\eta)-\eta\mathbf{x}_{1}\mathbf{\cdot x}_{3}-\eta\mathbf{x}%
_{2}\mathbf{\cdot x}_{3}>1$. If we take the $\mathbf{x}_{i}$ to be in the same
plane and $\frac{2\pi}{3}$ apart we get $\eta>0.8$.

The test of this violation is essentially the same as in the non-noisy case:
we take pairs from a source in the state (\ref{17}) and in each run we measure
the spin of the left particle along one of the directions $\mathbf{x}%
_{1}\mathbf{,x}_{2}$, or $\mathbf{x}_{3}$, and the spin of the right particle
along one of the same directions. The only difference is that from the
subensemble in which the same direction is chosen on both sides we can infer
the value of $\eta$, and then predict the size of the violation, and compare
it with the measured result.

Note that this is a worst case result, as are all the results that follow. We
have assumed that the rate of breakdown of symmetry between Alice and Bob is
the worst possible in the range given by (\ref{19}). In the average case the
amount of noise that can be tolerated is higher. If, for example, the symmetry
breaking in (\ref{19}) is zero on average then we can take any $\eta>\frac
{2}{3}$. Zero average symmetry breaking has been implicitly assumed in a
recent paper by Wildfeuer and Dowling \cite{20}, and the result is a dramatic
improvement on the known bound on $\eta$.

We can apply the same argument to the clique-web inequality (\ref{11})
assuming that the $X_{i}$'s correspond to measurements made by Alice and the
$Z_{j}$'s by Bob (which means reversing the directions of the $\mathbf{z}_{j}%
$). Taking into account that $p-q=2r+1$ and $\left\vert W_{p}^{r}\right\vert
=\frac{pq}{2}$ we get%
\begin{equation}
\eta>\frac{p}{(r+1)V_{p}^{r}+p-r-1}\label{22}%
\end{equation}
where $V_{p}^{r}$ is given in (\ref{12}). If we take the $(2k+1)$ bouquet in
(\ref{14}) and the limit $k\rightarrow\infty$ we get again $\eta>0.8$. It is
quite possible that a better result can be obtained by choosing other
arrangements of unit vectors.

We can apply the same consideration to the state $W$ guaranteed in Corrolary 3
to Tsirelson's theorem. Define%
\begin{equation}
\rho_{\eta}^{W}=\eta W+\frac{1-\eta\,}{d^{2}}I,\label{23}%
\end{equation}
where $d$ is the dimension of $\mathcal{H}$ in corrolary 3. Since
$tr[\rho_{\eta}^{W}(A_{i}\otimes B_{j})]=\eta\mathbf{x}_{i}\cdot\mathbf{x}%
_{j}$, and in particular $tr[\rho_{\eta}^{W}(A_{i}\otimes B_{i})]=\eta$, the
same argument can be repeated. Suppose that $b_{ij}$ is an $n\times n$ real
matrix normalized so that $\underset{X_{i}=\pm1}{\sup}\left(  \sum_{1\leq
i<j\leq n}b_{ij}\,X_{i}X_{j}\right)  =1$, and try to derive the best
substitute for (\ref{16}) in the presence of noise. In order to minimize the
noise choose a subset $J\subset\{1,2,...,n\}$ and associate the random
variables $\{X_{i}\ ;i\in J\}$ with Alice, and $\{Y_{j}\ ;j\in\overline{J}\}$
with Bob. In other words, on the right hand side of (\ref{16}) substitute%
\[
\sum_{1\leq i<j\leq n}b_{ij}\,X_{i}X_{j}\rightarrow\sum_{i<j,ij\in J}%
b_{ij}X_{i}X_{j}+\sum_{i\in J}\sum_{j\in\overline{J}}b_{ij}X_{i}Y_{j}%
+\sum_{i<j,ij\in\overline{J}}b_{ij}Y_{i}Y_{j}.
\]
Let $i<j$, if $(i,j)\in(J\times$ $\overline{J})\cup$ $(\overline{J}\times J)$
then substitute on the left hand side of (\ref{16}) $b_{ij}\mathbf{x}_{i}%
\cdot\mathbf{x}_{j}\rightarrow\eta b_{ij}\mathbf{x}_{i}\cdot\mathbf{x}_{j}$.
If $(i,j)\in(J\times J$ $)\cup$ $(\overline{J}\times\overline{J})$ then assume
worst case in (\ref{19}), and substitute in (\ref{16}) $b_{ij}\mathbf{x}%
_{i}\cdot\mathbf{x}_{j}\rightarrow\eta b_{ij}\mathbf{x}_{i}\cdot\mathbf{x}%
_{j}-\left\vert b_{ij}\right\vert (1-\eta)$. The accumulated noise will be
minimized when $J$ is chosen so the following value obtains
\begin{equation}
N\{b_{ij}\}=\underset{J}{\ \min}(\sum_{i<j,ij\in J}\left\vert b_{ij}%
\right\vert +\sum_{i<j,ij\in\overline{J}}\left\vert b_{ij}\right\vert
)=\underset{Z_{j}=\pm1}{\min}\frac{1}{2}\sum_{i\neq j}\left\vert
b_{ij}\right\vert (1+Z_{i}Z_{j}),\label{24}%
\end{equation}
(That is, we have to solve MAX CUT for the complete graph $K_{n}$, with
weights $\left\vert b_{ij}\right\vert $, and then subtract the outcome from
the total weight $\sum\left\vert b_{ij}\right\vert $). In all, the condition
for LHV model will be violated when%
\begin{equation}
\eta>\frac{N\{b_{ij}\}+1}{N\{b_{ij}\}+\left\vert \sum_{1\leq i<j\leq n}%
b_{ij}\,\mathbf{x}_{i}\cdot\mathbf{x}_{j}\right\vert }.\label{25}%
\end{equation}
The task is therefore to choose $b_{ij}$ and $\mathbf{x}_{1},\mathbf{x}%
_{2}...,\mathbf{x}_{n}\in\mathbb{R}^{n}$ such that the right hand side of
(\ref{25}) is minimal. It is known that $\eta>K_{G}^{-1}(n)$, where $K_{G}(n)$
are the Grothendieck constants in (\ref{9}). Indeed, for lower values of
$\eta$ all the bipartite Bell inequalities (for $\pm1$-valued, zero
expectation random variables) are satisfied by the quantum expectations. This
follows from (\ref{9}) and Tsirelson's theorem. Now, the satisfaction of these
inequalities is a sufficient condition for the existence of a LHV theory for
traceless observables (\cite{6}, see also the Appendix A). Hence, we get from
(\ref{25}) an interesting combinatorial inequality, which is proved on the
basis of physical considerations!

Note also that if $(b_{ij})$ is bipartite then $N\{b_{ij}\}=0$; hence, an
optimal choice of bipartite $n\times n$ matrices $(b_{ij})$ in (\ref{25}) will
give us $K_{G}^{-1}([\frac{n}{2}])$ on the right hand side. Therefore, as
$n\rightarrow\infty$ we get in (\ref{25}) the limit $K_{G}^{-1}$ as a lower
bound on the value of $\eta$ for all $n$.

Now, suppose that we take a fixed positive amount of noise, can we still get
an unbounded violation of Bell's inequality as $n\rightarrow\infty$? The
answer is yes, at least when the noise is less than $20\%$. To see that
consider the expression obtained when we substitute a value of $\eta$ in the
inequality. Put%
\[
V_{n}(\eta)=\eta\left\vert \sum_{1\leq i<j\leq n}b_{ij}\,\mathbf{x}_{i}%
\cdot\mathbf{x}_{j}\right\vert -(1-\eta)N\{b_{ij}\},
\]
where we choose $(b_{ij})$ that satisfy (\ref{16}). We know that if $\eta\leq
K_{G}^{-1}$ then $V_{n}(\eta)\leq1$, and the inequality is not violated. If
$\eta=1$ then $V_{n}(1)$ grows to infinity with $n$, and the violation is
unbounded. We also know from the two and three-dimensional cases that for
$\eta>0.8$ we get $V_{2}(\eta),V_{3}(\eta)>1$. Since the $n$ dimensional case
obviously includes the lower dimensional ones, we can find for each $n$ a real
number $\eta_{n}\leq0.8$ such that $V_{n}(\eta_{n})=1$. Hence, we get%
\[
V_{n}(0.9)=(0.9-\eta_{n})(\left\vert \sum_{1\leq i<j\leq n}b_{ij}%
\,\mathbf{x}_{i}\cdot\mathbf{x}_{j}\right\vert +N\{b_{ij}\})+V_{n}(\eta
_{n})\rightarrow\infty,
\]
The same analysis applies whenever the amount of noise is less than $20\%$.
Again, notice that this result concerns the worst possible symmetry breaking
in (\ref{19}).

\textbf{Acknowledgement:} This research is supported by the Israel Science
Foundation, grant 744/07.

\section{Appendix A: Polytopes and complexity}

Let $\mathbf{X=(}X_{1},X_{2},...,X_{n})$, $\mathbf{Y=(}Y_{1},Y_{2},...,Y_{m})$
be vectors with entries in $\{-1,1\}$. Denote by $\sigma_{ij}(\mathbf{X,Y}%
)=X_{i}Y_{j}$ and consider it as a vector in $\mathbb{R}^{nm}$ with
lexicographic order on the indices. The convex hull, in $\mathbb{R}^{nm}$, of
$\{\sigma_{ij}(\mathbf{X,Y});\ \mathbf{X}\in\{-1,1\}^{n},\mathbf{Y}%
\in\{-1,1\}^{m}\}$ is a polytope, call it $BELL(n,m)$. The face inequalities
for $BELL(n,m)$ have the form $\sum_{i,j}\alpha_{ij}X_{i}Y_{j}\leq\alpha$,
where $\alpha_{ij}$ and $\alpha$ are real numbers. The inequality is valid if
and only if it is satisfied by all the vertices $\sigma_{ij}(\mathbf{X,Y}%
)=X_{i}Y_{j}$ of $BELL(n,m)$. It represents a \emph{facet} if, in addition,
equality holds for a subset of the $\sigma_{ij}$'s which spans an affine
subspace of codimension one. For example the CHSH inequalities (\ref{3}) are
facet inequalities of $BELL(2,2)$, and all the non-trivial inequalities of
that polytope have that same shape.

A related structure is the polytope $BELL(n)$: Given $\mathbf{X=(}X_{1}%
,X_{2},...,X_{n})\in\{-1,1\}^{n}$ define $\sigma_{ij}(\mathbf{X})=X_{i}X_{j}$
for $1\leq i<j\leq n$ and consider $\sigma_{ij}(\mathbf{X})$ as a vector in
$\mathbb{R}^{\frac{1}{2}n(n-1)}$. \ $BELL(n)$ is the convex hull of
$\{\sigma_{ij}(\mathbf{X});\ \mathbf{X}\in\{-1,1\}^{n}\}$ in $\mathbb{R}%
^{\frac{1}{2}n(n-1)}$. For both $BELL(n,m)$ and $BELL(n)$ finding all the
inequalities is an impossible task (see below). In this paper we are using the
fact that more inequalities are known for $BELL(n)$ than for the general case
$BELL(n,m)$. Since the singlet state entails $X_{i}=-Y_{i}$ \ when Alice and
Bob are measuring in the same directions, we can use the known inequalities of
$BELL(n)$. Note also the following relation between the polytopes: Let $u\in$
$BELL(n)$ be a $\frac{1}{2}n(n-1)$- dimensional vector, define $v\in$
$\mathbb{R}^{n^{2}}$ by $v_{ij}=v_{ji}=u_{ij}$ when $1\leq i<j\leq n$ and
$v_{ii}=1$, then $v\in$ $BELL(n,n)$. This means that every valid inequality
for $BELL(n,n)$ can be collapsed to a valid inequality for $BELL(n)$.
Moreover, if deciding membership in $BELL(n)$ is a computationally hard task,
then so is deciding membership in $BELL(n,n)$.

Similarly, let $\mathbf{a}=(a_{1},a_{2},...,a_{n})\in\{0,1\}^{n}$ for $1\leq
i<j\leq n$. Denote $\delta_{ij}(\mathbf{a})=a_{i}\oplus a_{j}=a_{i}%
+a_{j}-2a_{i}a_{j}$ and consider it as a vector in $\mathbb{R}^{\frac{1}%
{2}n(n-1)}$. The cut polytope $CUT(n)$ is the convex hull of $\{\delta
_{ij}(\mathbf{a});\ \mathbf{a}\in\{0,1\}^{n}\}$. The relations between
$CUT(n)$ and $BELL(n)$ are not hard to determine. Since for all $1\leq i<j\leq
n$
\begin{equation}
\delta_{ij}(\mathbf{a})=a_{i}\oplus a_{j}=\frac{1-X_{i}X_{j}}{2}%
=\frac{1-\sigma_{ij}(\mathbf{X})}{2}\quad where\ X_{i}=2a_{i}-1 \label{26}%
\end{equation}
we conclude that $(v_{ij})\in$ $BELL(n)$ if and only if $(%
\frac12
(1-v_{ij}))\in CUT(n)$. There is extensive work on the facets of the cut
polytope \cite{18}. Each facet inequality can be readily transferred to
$BELL(n)$.

Finally, let $\mathbf{b}=(b_{1},b_{2},...,b_{n})\in\{0,1\}^{n}$, for $1\leq
i\leq j\leq n$ denote $\pi_{ij}(\mathbf{b})=b_{i}b_{j}$ and consider it as a
vector in $\mathbb{R}^{\frac{1}{2}n(n+1)}$ with lexicographic order on the
indices. The convex hull in $\mathbb{R}^{\frac{1}{2}n(n+1)}$ of \ $\{\pi
_{ij}(\mathbf{b});\ \mathbf{b}\in\{0,1\}^{n}\}$ is called correlation
polytope, and denoted by $COR(n)$. (Note that $COR(n)$ has dimension $\frac
{1}{2}n(n+1)$ while $CUT(n)$ and $BELL(n)$ have dimension $\frac{1}{2}%
n(n-1)$). For these polytope we have: \cite{18, 21}

\begin{theorem}
(\textbf{a}) Let $(p_{ij})\in\mathbb{R}^{\frac{1}{2}n(n+1)}$ then $(p_{ij})\in
COR(n)$ if and only if there is a probability space $(X,\Sigma,\mu)$ and
events $E_{1},E_{2},...,E_{n}\in\Sigma$ such that $p_{ij}=\mu(E_{i}E_{j})$ for
$1\leq i\leq j\leq n$ . \ (\textbf{b}) Let $(c_{ij})\in\mathbb{R}^{\frac{1}%
{2}n(n-1)}$ then $(c_{ij})\in CUT(n)$ if and only if there is a probability
space $(X,\Sigma,\mu)$ and events $E_{1},E_{2},...,E_{n}\in\Sigma$ such that
$c_{ij}=\mu(E_{i}\bigtriangleup E_{j})=\mu\lbrack(E_{i}\setminus E_{j}%
)\cup(E_{j}\setminus E_{i})]=$ $\mu(E_{i})+\mu(E_{j})-2\mu(E_{i}E_{j})$ for
$1\leq i<j\leq n$.
\end{theorem}

For both the cut polytope and the correlation polytope we can easily extend
the definition to the bipartite cases $CUT(n,m)$ and $COR(n,m)$ in a
straightforward way. The relations between the one-sided polytopes and their
bipartite versions are similar to that of the Bell polytope..

For the correlation polytope we have the following complexity results
\cite{15} which can easily be transferred to the cut polytope and the Bell polytope.

{\large 1. }Deciding whether a given rational $(p_{ij})\in\mathbb{R}^{\frac
{1}{2}n(n+1)}$ is an element of $COR(n)$ is an $NP$-complete problem . (This
remains valid when $p_{ii}=\frac{1}{2}$ for all $1\leq i\leq n$.)

{\large 2. }Deciding whether a given inequality is \emph{not} valid for
$COR(n)$ is an $NP$-complete problem.

All this means that unless $NP=P$ (or at least $NP=coNP$) deriving all the
inequalities for any of these polytopes is a computationally impossible task
for large $n$. This does not prevent us from deriving special cases or even
infinite families of inequalities.

\subsection{Appendix B: The Clique-Web inequalities of $CUT(n)$}

A graph $G=(V,E)$ consists of a set of vertices $V_{n}=\{1,2,...,n\}$ and a
set of edges $E$, which are just (unordered) pairs of vertices. If the set of
vertices $V_{n}$ has been fixed we shall often speak loosely on `the graph
$E$' mentioning only the edges. The set of all pairs on $n$ vertices is called
the \emph{complete} graph and denoted by $K_{n}$.

Let $S\subset V_{n}=\{1,2,...,n\}$ be a non empty subset of vertices. Denote
by $\kappa(S)=\{\{i,j\};\ i\neq j,\;i,j\in S\}$. If $\kappa(S)\subset E$ then
$\kappa(S)$ is called a clique in the graph $G=(V_{n},E)$. Also, define a
graph $\delta(S)=\{\{i,j\};\ i\in S,j\notin S\ or\ i\notin S,j\in S\}$. The
graph $\delta(S)$ is called \emph{a} \emph{cut} (or a cut in $K_{n}$). Denote
by $\mathbf{a=}(a_{1},a_{2},...,a_{n})\in\{0,1\}^{n}$ the indicator function
of $S$, so that $a_{i}=1$ for $i\in S$ and $a_{i}=0$ otherwise. Then
$\{i,j\}\in$ $\delta(S)$ if and only if $a_{i}\oplus a_{j}=a_{i}+a_{j}%
-2a_{i}a_{j}=1$. Hence, the vertices of the cut polytope $CUT(n)$, that is
$\delta_{ij}(\mathbf{x})=a_{i}\oplus a_{j}$, are the indicator functions of
the cuts $\delta(S)$.

The Inequalities that we have considered follow from a precise
characterization of $\delta(S)\cap E$ for a graph related to $W_{p}^{r}$.
Recall that if $p,q$, and, $r$ ne three integers such that $q\geq2$ and
$p-q=2r+1$ the \emph{web }$W_{p}^{r}$ is the graph whose set of vertices is
$V_{p}=\{1,2,...,p\}$ and set of edges is
$\{i,i+r+1\},...,\{i,i+r+q\}\ ;\ i=1,2,...,p,\ $%
addition\ is$\ \operatorname{mod}p.$ The \emph{antiweb }$AW_{p}^{r}$ \ is the
complement in $K_{p}$ of the web $W_{p}^{r}$.

For these graphs Alon \cite{17} proved the following

\begin{theorem}
Let $p$, $r$ be integers such that $p\geq2r+3$, $r\geq1$. Let $S\subset
\{1,2,...,p\}$ and assume that $\left|  S\right|  =s$.

1. If $s\leq r$, then $\left|  \delta S\cap AW_{p}^{r}\right|  $ $\geq
s(2r+1-s)$, with equality if and only if $\kappa(S)$ is a clique in
$AW_{p}^{r}$.

2. If $r+1\leq s\leq\frac{p}{2}$, then $\left\vert \delta S\cap AW_{p}%
^{r}\right\vert \geq r(r+1)$ with equality if and only if $S$ is an interval
in $\{1,2,...,p\}$, that is, it has the form $S=\{i,i+1,i+2,...,i+s-1\}$ for
some $i$, $1\leq i\leq p$ (addition is $\operatorname{mod}p$.)
\end{theorem}

As an outcome of Alon's theorem Deza and Laurent proved the following
inequality for $CUT(p+q)$ where ($q=p-2r-1$):%

\begin{equation}
\sum_{\{i,j\}\in W_{p}^{r}}a_{i}\oplus a_{j}+\sum_{1\leq i<j\leq q}b_{i}\oplus
b_{j}-\sum_{i=1}^{p}\sum_{j=1}^{q}a_{i}\oplus b_{j}\leq0 \label{27}%
\end{equation}
(See \cite{18} for details). Here $\mathbf{a=}(a_{1},a_{2},...,a_{p}%
)\in\{0,1\}^{p}$ and $\mathbf{b=}(b_{1},b_{2},...,b_{q})\in\{0,1\}^{q}$ . If
we substitute $X_{i}=2a_{i}-1$ and $Z_{j}=2b_{j}-1$ and use the identity
(\ref{26}), we get the inequality (\ref{11}) for $BELL(p+q)$.

\bigskip

\bigskip

\end{document}